\documentclass[aps,prl,twocolumn,superscriptaddress,groupedaddress]{revtex4}
\usepackage{graphicx}  
\usepackage{dcolumn}   
\usepackage{bm}        
\usepackage{amssymb}   
\usepackage{xcolor}
\usepackage{color}
\usepackage{mathtools}
\usepackage{amssymb}
\usepackage{amsmath}
\usepackage{braket}
\usepackage{bbold}
\usepackage{soul}
\usepackage{relsize}

\begin{document}

\title{Topological rejection of noise by quantum skyrmions}
\author{Pedro Ornelas}
\affiliation{School of Physics, University of the Witwatersrand, Private Bag 3, Wits 2050, South Africa}
\author{Isaac Nape}
\affiliation{School of Physics, University of the Witwatersrand, Private Bag 3, Wits 2050, South Africa}
\author{Robert de Mello Koch}
\affiliation{School of Science, Huzhou University, Huzhou 313000, China}
\affiliation{Mandelstam Institute for Theoretical Physics, School of Physics,University of the Witwatersrand, Private Bag 3, Wits 2050, South Africa}
\author{Andrew Forbes\footnote{Corresponding author. e-mail: andrew.forbes@wits.ac.za}}
\affiliation{School of Physics, University of the Witwatersrand, Private Bag 3, Wits 2050, South Africa}

\begin{abstract}

\noindent \textbf{An open challenge in the context of quantum information processing and communication is improving the robustness of quantum information to environmental contributions of noise, a severe hindrance in real-world scenarios. Here, we show that quantum skyrmions and their nonlocal topological observables remain resilient to noise even as typical entanglement witnesses and measures of the state decay. This allows us to introduce the notion of digitization of quantum information based on our new discrete topological quantum observables, foregoing the need for robustness of entanglement. We compliment our experiments with a full theoretical treatment that unlocks the quantum mechanisms behind the topological behaviour, explaining why the topology leads to robustness. Our approach holds exciting promise for intrinsic quantum information resilience through topology, highly applicable to real-world systems such as global quantum networks and noisy quantum computers.}
\end{abstract}

\maketitle
\section{Introduction}
\noindent Quantum entanglement is an essential and viable resource for future quantum technologies. It has proven its worth in applications as diverse as secure quantum communication \cite{gisin2002quantum, scarani2009security, ma2007quantum}, even over large distances in satellite based networks \cite{liao2017satellite,bedington2017progress}, quantum computing \cite{thomas2022efficient, raussendorf2001one,briegel2009measurement}, quantum imaging of complex objects \cite{pepe2016correlation,sephton2023revealing}, quantum metrology \cite{giovannetti2011advances} and lithography \cite{boto2000quantum, kok2001quantum}.  However, it is well known that noise deteriorates entanglement, for instance, due to noisy detectors (shot and thermal noise), stray light (e.g., daylight), depolarising channels, lost photons to name but a few, all examples of generic (isotropic) noise that results in the decay of quantum information in realistic quantum scenarios \cite{nielsen2001quantum, ecker2019overcoming, zhu2021high}.  The consequence is a diminished ability to certify, identify and exploit the full benefits of quantum entanglement for information processing \cite{kumar2003effect, nielsen2001quantum, lloyd1997capacity, liang2013quantum}. High dimensional ($d>2$ dimensions) entangled states are able to increase the noise threshold to which the usual entanglement witnesses certify entanglement 
\cite{ecker2019overcoming,peterfreund2021multidimensional,almeida2007noise,qu2022robust,tsokeng2018dynamics,qu2022retrieving} but comes with the cost of limited state purity \cite{zhu2021high} and complex quantum state analysis \cite{nape2023quantum}. Although some mitigation techniques exist such as real-time channel analysis \cite{ndagano2017characterizing} and entanglement purification \cite{yan2023advances}, real-world noise correction remains challenging, prohibiting noise-free communication when information is distributed by entanglement.  \\

An emerging trend is to exploit quantum topological photonics \cite{yan2021quantum}, which has already realised stable quantum emitters \cite{mehrabad2020chiral,dai2022topologically,mittal2018topological}, robust transport of entanglement through quantum circuits \cite{barik2018topological,blanco2018topological} and entanglement storage \cite{parmee2022optical} but here the topology is not in the quantum state but in topologically structured matter. Recently the notion of optical topology has emerged, specifically optical skyrmions \cite{shen2024optical}, with the potential to embed measurable topological properties into the non-separable spatial and polarization degrees of freedom of classical \cite{gao2020paraxial, shen2021topological,shen2022generation, singh2023synthetic,sugic2021particle} and quantum \cite{ornelas2024non} photonic states.  A motivation for the interest in skyrmions is their potential to act as resilient information carriers in the presence of noise, but numerical studies have returned mixed outcomes \cite{liu2022disorder} and to date no experimental evidence for realistic noise channels exists.\\

\begin{figure}[t!]
    \includegraphics[width=1\linewidth]
{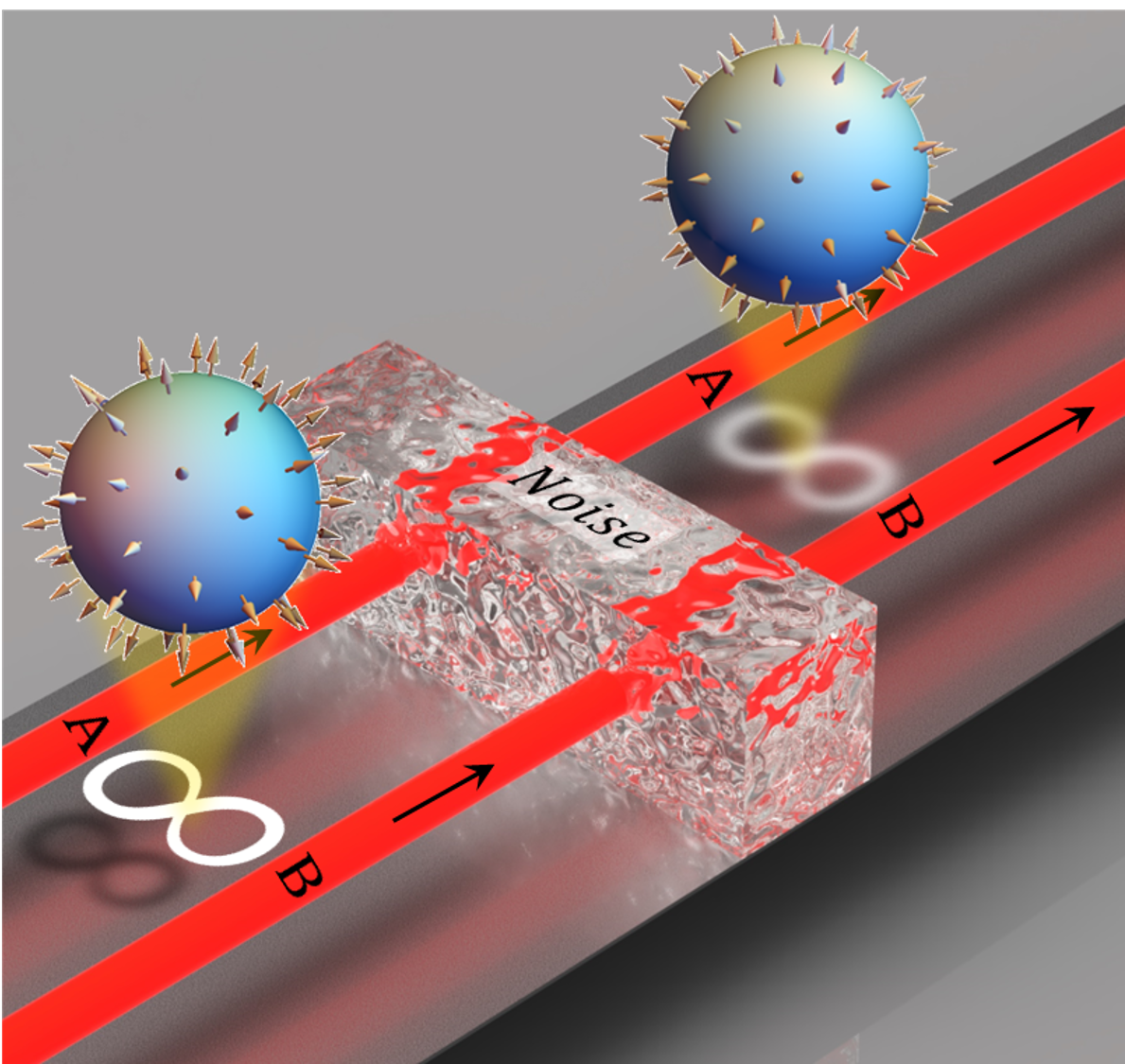}
\caption{\textbf{Quantum Skyrmions through noise.} Photons A and B form an entangled state characterized by a non-local quantum skyrmion, shown as a stereographic projection of vectors on a sphere. Passing such a state through noise diminishes the quality of the entangled state, but leaves the topological observable unaltered so long as some entanglement persists.}\label{Figure 1} 
\end{figure}

Here we show both theoretically and experimentally that quantum skyrmions and their topological quantum observables remain resilient to noise even as typical entanglement witnesses and quantum observables of the state decay. 
To realise this we create a nonlocal skyrmionic topology as a shared property of two entangled photons and pass the quantum topology through noise, as illustrated in Fig.~\ref{Figure 1}, which we simulate by a generic (isotropic) noise model, a standard approach to test quantum states in noise \cite{horodecki2000limits, horodecki1999general, horodecki2009quantum}. We provide an intuitive explanation, supported by a rigorous theoretical treatment, as to why such noise can be considered a smooth deformation of the state - an operation that does not modify its topology. We present experimental results across various topologies and for a wide range of noise levels, confirming the topological invariance despite the decay of typical entanglement witnesses. Our work opens a new path to quantum information processing and communication in noisy quantum systems and channels without the need for compensation or mitigation strategies.

\begin{figure*}[t!]
    \includegraphics[width=1\linewidth]
{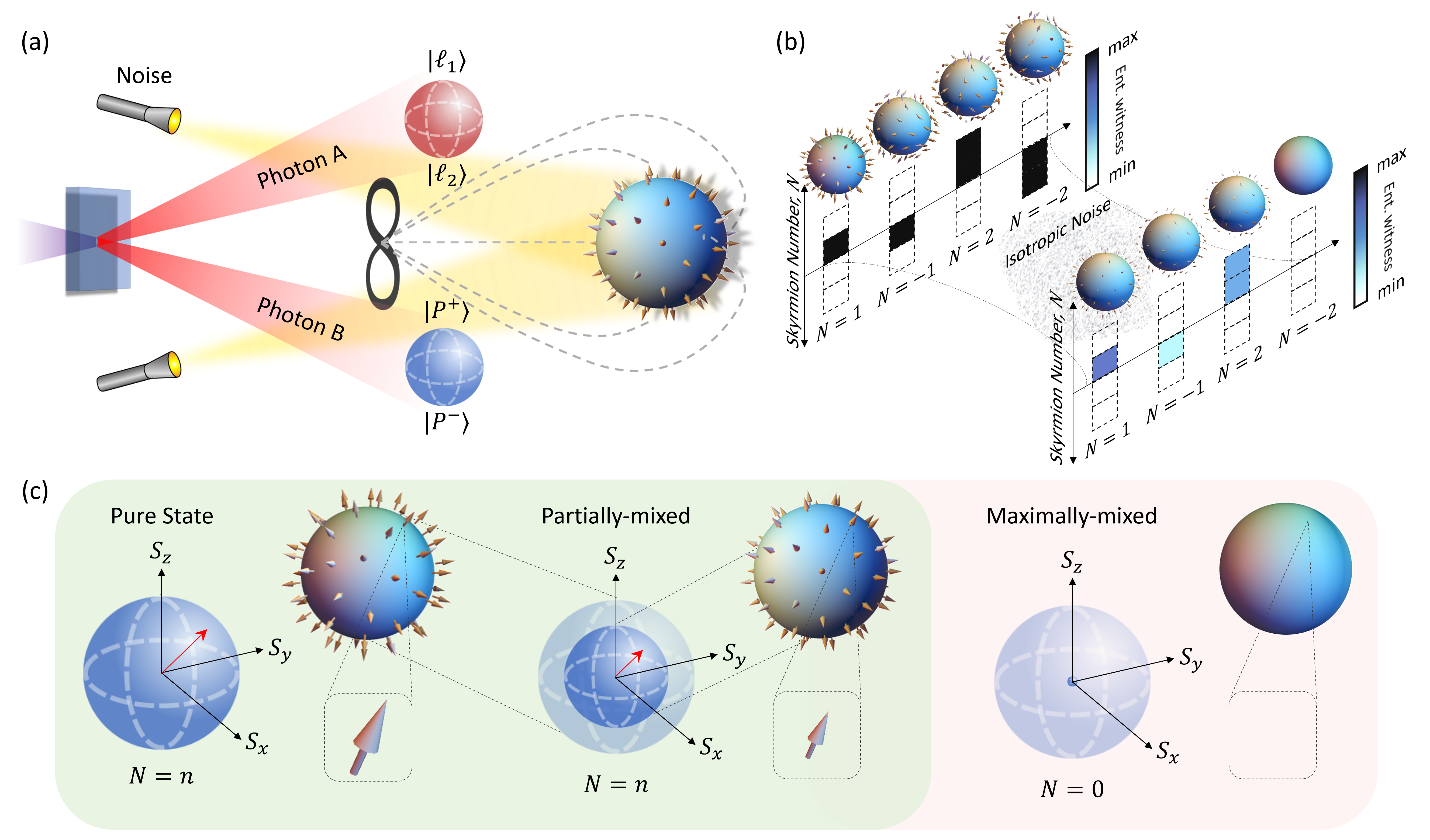}
\caption{\textbf{Topological rejection of noise.} (a) We create a bi-photon entangled state, with photon A in the spatial degree of freedom and photon B in the polarization degree of freedom, for a non-local skyrmionic topology as a shared emergent property of the two photons, shown as a stereographic projection. (b) The Skyrmion number is a digitized observable, shown here for integer values from $N = -2$ to $N = 2$, remaining intact along with the topology after the isotropic noise even while the entanglement decreases, except in the case when the entanglement reaches its minimum, indicating a non-entangled state.  
(c) In the geometric picture of photon B's Hilbert space as expressed by Stokes vectors ($S_x,S_y,S_z$), the change of state purity due to noise is a sphere of diminishing size, which for partially mixed states can always be renormalized while leaving the topology unaltered. Once the state is maximally-mixed, the Hilbert space of photon B has zero radius, no entanglement, and thus only a trivial topology ($N = 0$).
    \label{Figure 2} }
\end{figure*}

\section{Results}
\noindent \textbf{Quantum skyrmions.} As our aim is to study quantum skyrmionic topology in the presence of noise, we begin by constructing a bi-photon entangled state of the form $\ket{\Psi} = \frac{1}{\sqrt{2}} \left( \ket{\ell_1}_A \ket{P_1}_B + e^{i\delta}\ket{\ell_2}_A \ket{P_2}_B \right)$, as depicted graphically in Fig.~\ref{Figure 2} (a). Here $\ell_1$ and $\ell_2$ denote orbital angular momentum (OAM) of $\ell_1 \hbar$ and $\ell_2 \hbar$ per photon, respectively, $\ket{P_1}, \ket{P_2}$ are orthogonal polarization states and $\delta$ allows for a relative phase between the two components of the state vector. When $|\ell_1| \neq |\ell_2|$ the bi-photon correlations form the desired skyrmionic mapping from $\mathcal{R}^2 \to \mathcal{S}^2$, so that the entanglement defines a topology \cite{ornelas2024non}. The topological invariant, the Skyrmion number ($N$), characterizes the number of times $\mathcal{R}^2$ (stereographically projected Hilbert space of photon A) wraps $\mathcal{S}^2$ (Hilbert space of photon B), i.e., the number of times photon B's collapsed states wrap its own Hilbert space after completing a full set of spatial position measurements on photon A. This mapping is evident when expressing the OAM states of photon A in the position basis, using $|\ell\rangle = \int_{\mathcal{R}^2} |\text{LG}_\ell \left( \mathbf{r} \right)| e^{i\ell\phi} |\mathbf{r}\rangle \ d^2r$ where $\text{LG}_\ell \left( \mathbf{r} \right)$ are the Laguerre-Gaussian fields and $|\mathbf{r}\rangle$ are position states corresponding to coordinates vectors $\mathbf{r}_A \in \mathcal{R}^2$. Ignoring global phase factors, this recasts our state to
   \begin{align}
    \ket{\Psi} = \int_{ \mathcal{R}^2 }  \ket{\mathbf{r}}_A \left( a(\mathbf{r}_A) \ket{P_1}_B + b(\mathbf{r}_A) \ket{P_2}_B \right) \ d^2r_A,  \label{Eq:SpatialquantumSkyrmion}
    \end{align}
    where the coefficients $a(\mathbf{r}_A) \equiv |\text{LG}_{\ell_{1}} \left( \mathbf{r}_A \right)|$ and $b(\mathbf{r}_A) \equiv \exp(i(\Delta\ell \phi_A + \delta)) |\text{LG}_{\ell_{2}} \left( \mathbf{r}_A \right)|$ are chosen to be normalized according to $|a(\mathbf{r}_A)|^2 + |b(\mathbf{r}_A)|^2 = 1$ for all $\mathbf{r}_A \in \mathcal{R}^2$ and $\Delta\ell = \ell_2-\ell_1$. The Skyrmion number of such a state is then calculated from \cite{gao2020paraxial}
    \begin{equation}
    N = \frac{1}{4\pi}\int_{\mathcal{R}^2} \epsilon_{pqr} S_p \frac{\partial S_q}{\partial x} \frac{\partial S_r}{\partial y} dx dy \, ,
    \label{skyrme}
    \end{equation}
     \noindent where $\epsilon_{pqr}$ is the Levi-Civita symbol and it is assumed that the sphere has been normalized such that $\Sigma_{i=1}^3 S_i^2 = 1$, i.e., a mapping to the unit sphere.  
To find the necessary quantum Stokes parameters, $S_i$, one computes the expectation values of the Pauli matrices, calculated by taking the diagonal matrix element at position $\mathbf{r}$ for photon A and the partial trace over photon B, such that $S_i = \text{Tr}_B(\langle |\mathbf{r}\rangle_A \langle \mathbf{r}|_A \otimes \sigma_{B,i}\rangle) = \text{Tr}_B \left(\sigma_{B,i}\,\,{}_A\langle\mathbf{r}|\Psi\rangle \langle \Psi|\mathbf{r}\rangle_A \right)$. 

\vspace{0.2cm}
\noindent \textbf{Topological protection.}  A common way to simulate real-world noisy conditions is to use a generic (isotropic) noise model, by mixing the initially pure state with a maximally mixed state \cite{dur1999quantum}, covering a wide range of random noise sources e.g., from the source (multi-photon events), the channel (background and lost photons), and the detector (dark counts).  To keep the conclusions general we will use the purity of the state as our measure of the noise, and not knowledge of the noise source itself, i.e., an agnostic measure of the noise impact.  Without loss of generality we restrict ourselves to the original two dimensional OAM subspace of the Hilbert space of photon A so that the density matrix can be written as
\begin{equation} 
\rho = p\,|\Psi\rangle\langle\Psi| + \frac{1-p}{4} \mathbb{1}_{4}, 
        \label{eq: PartialMixed}
\end{equation}
\noindent and $\mathbb{1}_4$ is the $4 \! \times  \! 4$ identity matrix on the tensor product of the Hilbert space of photon B and the subspace of photon A that we restrict to. See the supplementary information for an explicit account. The purity of this state is given by $\gamma =\text{Tr}(\rho^2)= p^2 + \frac{1-p^2}{4}$ with $p \in [0,1]$, e.g., a pure state ($|\Psi\rangle \langle \Psi|$) for $p=1$ and a maximally mixed state ($\frac{1}{4}\mathbb{1}_{4}$) for $p=0$.  To calculate the Skyrmion number we must determine the effect of the noise on the quantum Stokes parameters, whose action is to produce new parameters, $S_i'$.  

    The proposed noise rejection mechanism can be understood if we consider the spectral decomposition of the Pauli observables, given by $\sigma_{B,i}= \lambda^+_iP^+_i + \lambda^-_iP^-_i$ where $P_i^{\pm} = |\lambda^{\pm}_i\rangle\langle\lambda^{\pm}_i|$ for positive and negative eigenvalues $\lambda^{\pm}=\pm 1$, thus 
    \begin{eqnarray}
        S_i'&=&\text{Tr}_B \left(P^+_i \left[p \,\, {}_A\langle\mathbf{r}|\Psi\rangle\langle\Psi|\mathbf{r}\rangle_A + \frac{1-p}{2}\mathbb{1_{2}} \right]\right)\cr\cr\cr
        &&-\text{Tr}_B\left(P^-_i \left[p\,\,{}_A\langle\mathbf{r}|\Psi\rangle\langle\Psi|\mathbf{r}\rangle_A + \frac{1-p}{2}\mathbb{1_{2}} \right]\right),
    \end{eqnarray}

    \noindent which exposes a subtle point: the noise rejection mechanism stems from subtracting measured values of the projections $P^{\pm}_i$, which each contain the same noise contribution, $\frac{1-p}{2}\text{Tr}_B\left(P^{\pm}_i \right)=\frac{1-p}{2}$. This invariance is depicted in Fig.~\ref{Figure 2} (b) where the Skyrmion number remains constant in the presence of noise despite the decay of the initial state quantified by a decrease in typical entanglement witnesses.  Normalizing each of the $S_i'$ parameters by $S_0^{\prime 2} = \text{Tr}_B \left({}_A\langle\mathbf{r}|\rho |\mathbf{r}\rangle_A \right)$ reveals the smooth deformation that takes place under the addition of this noise to the system. Without this normalization it can be shown that $\sum_i^3 S_i^{\prime 2} = \frac{4p^2}{(p+1)^2} \leq 1$
which reveals that photon B's state is mapped to a state on a shell in the interior of its Hilbert space (a unit two sphere), with a radius dictated by the purity of the state, as depicted in Fig.~\ref{Figure 2} (c). This does not change the topology: $\mathcal{R}^2$ still maps to a sphere $\mathcal{S}^2$, albeit one of a smaller radius, with the same Skyrmion number. When the state is maximally mixed, $\gamma=\frac{1}{4}$, the calculated Skyrmion number vanishes as the radius of the shell has reduced to zero. In this singular limit $\mathcal{R}^2$ maps to a single point, depicted in Fig.~\ref{Figure 2} (c), consistent with the fact that when there is no entanglement, there can be no non-trivial topology \cite{ornelas2024non}. The topology remains robust to the noise so long as \textit{some} entanglement persists.

\begin{figure*}[t!]
    \includegraphics[width=1\linewidth]
{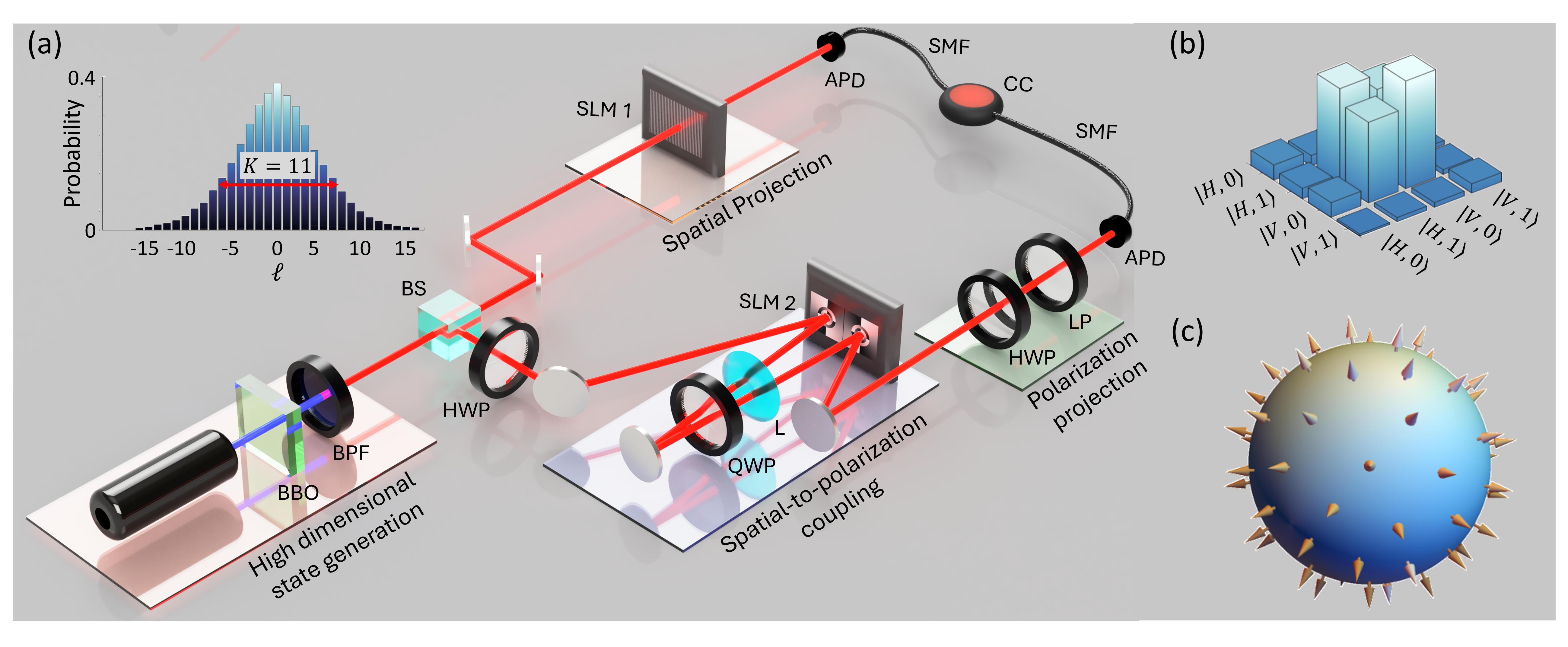}
\caption{\textbf{Experimental generation of Quantum Skyrmions.} (a) Experimental setup for the generation, preparation and detection of non-local skyrmionic quantum states, with details given in the main text and Supplementary Information. The desired hybrid states are prepared via a spatial-to-polarization coupling (SPC) technique. A spiral bandwidth measurement (shown as an inset) reveals the available OAM modes in the initial high-dimensional entangled state.  
To confirm the system performance we show (b) the reconstructed experimental density matrix and non-local topology for the ideal example state, $\ket{\Psi} = \frac{1}{\sqrt{2}} \left( \ket{0}_A\ket{H}_B + \ket{1}_A \ket{V}_B \right)$. Abbreviations: Barium Borate (BBO), band-pass filter (BPF), 50:50 beam splitter (BS), lens (L), spatial light modulator (SLM), avalanche photodiode (APD), single mode fibre (SMF), coincidence counter (CC), half-wave plate (HWP), quarter-wave plate (QWP), linear polarizer (LP). \label{Figure 3} }
\end{figure*}

    \begin{figure}
        \includegraphics[width=\linewidth]{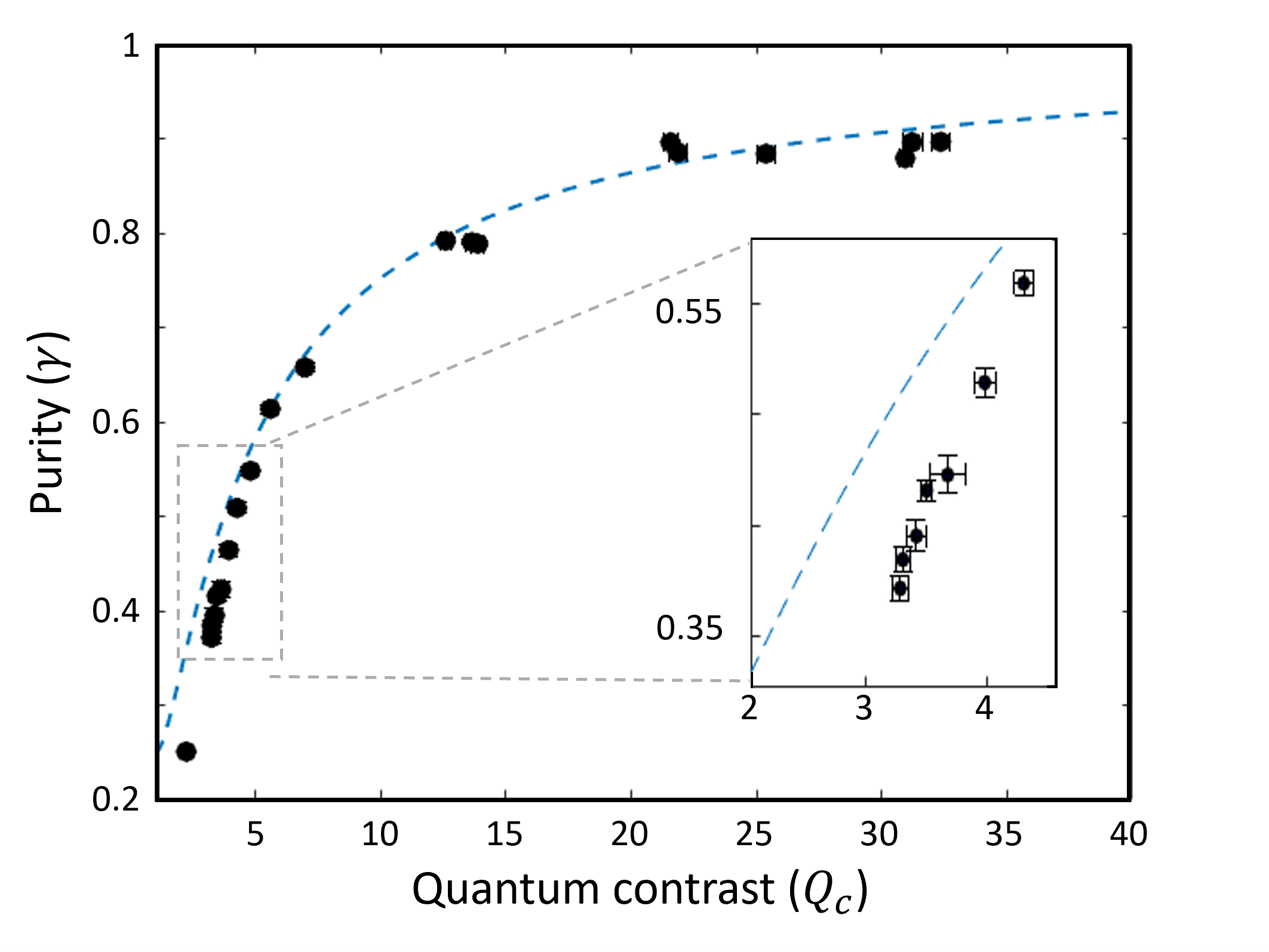}
        \caption{\textbf{Noise characterization.} Purity as a function of quantum contrast, with theory (dashed lines) and experiment (points) in good agreement.  The inset is used to show the error bars from the experiment. We use quantum contrast as our measure of the noise impact as it is agnostic to the noise source.
        \label{Figure 4} }
    \end{figure}


  \vspace{0.2cm}
  \noindent \textbf{Experiment.}  To verify the concept we use the experiment depicted in Fig.~\ref{Figure 3}(a), with full details given in the SI. Entangled OAM biphotons of wavelength 710 nm were generated through spontaneous parametric down-conversion (SPDC) in a 3 mm long Type-I Barium Borate (BBO) nonlinear crystal pumped by a 355 nm wavelength collimated Gaussian beam. A spiral bandwidth measurement (top left inset) was performed from which a Schmidt number of $K=11$ was calculated \cite{nape2023quantum}. This allows for hybrid state generation with OAM order as high as $\ell=\pm5$. To achieve our topologically non-trivial states we used the spatial-to-polarization coupling (SPC) state preparation stage, exchanging the OAM-OAM correlations between photons A and B for OAM-Polarization correlations.  To do so, Photon A was left unaltered while photon B's spatial information was coupled to its polarization using a double bounce off a spatial light modulator (SLM B). The incoming horizontally polarized photons were modulated in the first pass and the vertically polarized photons in the second (by splitting the SLM screen in two parts), with the quarter-wave plate (QWP) serving to accommodate the fact that the SLM can only modulate horizontally polarized light.  For non-trivial topologies the holograms displayed on SLM B were set such that $|\ell_1| - |\ell_2| \neq 0$. In the detection stage, photon A was directed to a spatial light modulator (SLM A) for spatial projective measurements while photon B was passed through a HWP orientated to $45^{\circ}$ and a linear polarizer orientated at $90^{\circ}$ for polarization projective measurements. Both photons were collected by single-mode fibres (SMFs) coupled to an avalanche photon detector (APD) and measured in coincidence. Six projective measurements on each photon allowed for an overcomplete quantum state tomography (QST) with 36 entries, which was used to reconstruct the quantum state (with and without noise), with full details given in the SI. Initially, the experiment was run without noise to certify the creation and detection of the example state $\ket{\Psi} = \frac{1}{\sqrt{2}} \left( \ket{0}_A\ket{H}_B + \ket{1}_A\ket{V}_B\right)$. Fig.~\ref{Figure 3} (b) shows the reconstructed density matrix and (c) non-local topology of the state. The data is in agreement with the results expected from the desired pure state, with a purity of $\gamma = 0.80$ and Skyrmion number $N=0.96 \pm 0.01$ (theory: $N=1$).
  To introduce noise we used an incoherent light source of variable intensity, which is known to be a good approximation to isotropic noise \cite{ecker2019overcoming}. The contribution of the white light source to the final state was determined from the average quantum contrast, $Q_c$, which estimated the signal-to-noise (SNR) ratio in our system (See SI for full details), reaching values as high as 32.3 and as low as 2.24 for low and high levels of noise, respectively. This impacts directly on the purity of the state following  $ \gamma = \frac{1}{4}\left[3\left(\frac{Q_c-1}{Q_c+1}\right)^2+1\right]$, which we used as a measure of the impact of the noise, ranging from our target pure state (no noise, $Q_c \gg 1$) to a maximally mixed (separable) state (high noise, $Q_c \approx 1$).  In Fig.~\ref{Figure 4} the experimentally obtained purity is plotted against average quantum contrast and is in excellent agreement with theory. This demonstrates that precise control over the experimental purity was achieved through control over the quantum contrast of the system.  

\begin{figure*}
        \includegraphics[width=\linewidth]{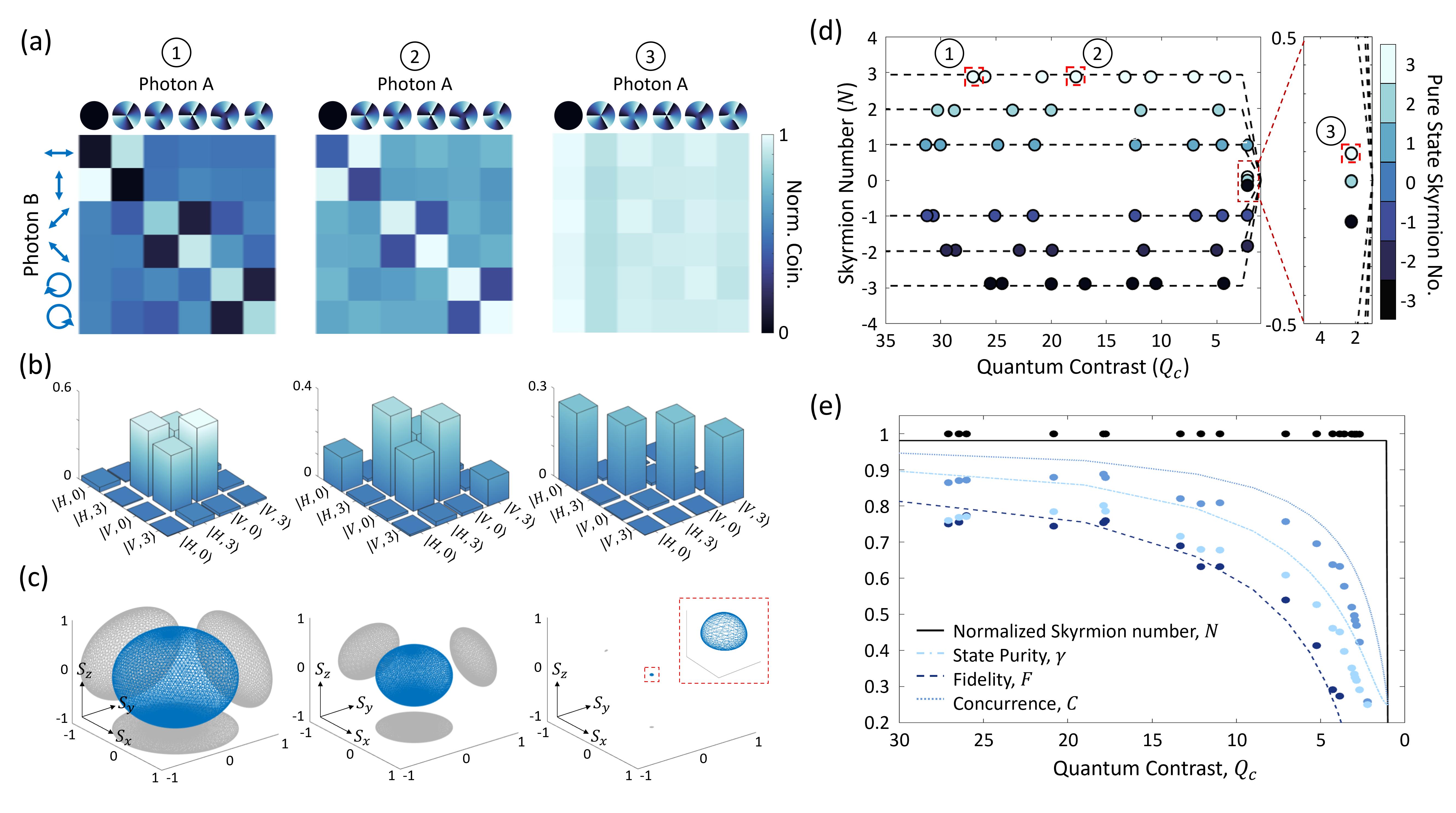}
        \caption{\textbf{Experimental verification of topological resilience.} (a) QST results for the example state $\ket{\Psi} = \frac{1}{\sqrt{2}} \left( \ket{0}_A\ket{H}_B + \ket{3}_A\ket{V}_B\right)$ in the presence of varying noise levels from pure (point 1, $\gamma = 0.80$) to partially-mixed (point 2, $\gamma = 0.45$) and then maximally-mixed (point 3, $\gamma = 0.25$). (b) The experimental density matrices extracted from the QST of each state. (c) The quantum Stokes vector, $\vec{S}(x,y) = (S_x, S_y, S_z)^T$, derived from each density matrix. (d) Skyrmion number as a function of quantum contrast, with experiment (points) in excellent agreement with theory (dashed lines). The collapse of the topology when the maximally-mixed state limit ($Q_c=1$) is reached is shown in the zoomed-in inset. (e) While the Skyrmion number is stable to noise, typical entanglement witnesses such as concurrence, fidelity and purity are not, as shown experimentally (points) and theoretically (lines).
        \label{Figure 5}}
    \end{figure*}

The experiment was run for seven different initial states with varying topology, $N\in \{-3,-2,-1,1,2,3\}$, each for a wide range of isotropic noise levels, and in each case the quantum Stokes parameters inferred from a quantum state tomography. Example data is shown in Fig.~\ref{Figure 5} (a-c)  for the states $\rho = p\ket{\Psi}\bra{\Psi} + \frac{1-p}{4}\mathbb{1}_4$ where $\ket{\Psi} = \frac{1}{\sqrt{2}} \left( \ket{0}_A\ket{H}_B + \ket{3}_A\ket{V}_B\right)$ in the presence of varying isotropic noise levels from pure (point 1, $\gamma = 0.80$) to partially-mixed (point 2, $\gamma = 0.45$) and then maximally mixed (point 3, $\gamma = 0.25$). As expected the QST data shown in Fig.~\ref{Figure 5}(a) reveals a characteristic saturation of each projection measurement with an increase in isotropic noise. This is accompanied by a clear transition of the density matrix describing a pure state (entangled) to that describing a maximally-mixed state (no entanglement) as shown in Fig.~\ref{Figure 5}(b). The quantum Stokes vector, $\vec{S}(x,y) = (S_x, S_y, S_z)^T$, derived from each density matrix is plotted in Fig.~\ref{Figure 5}(c). As expected, the magnitude of the quantum Stokes vector diminishes with decreasing purity (increasing noise), until it collapses to a point with zero magnitude. The Skyrmion number for these three states (1-3) was calculated to be $N=2.99 \pm 0.01$, $N=2.99 \pm 0.01$ and $N=0.01 \pm 0.01$, respectively, revealing that the skyrmion number remains intact until the entanglement vanishes altogether.

Fig.~\ref{Figure 5}(d) shows data for the rest of the entangled states considered. The theoretical (dashed lines) and experimental (points) results for the Skyrmion number as a function of quantum contrast are in excellent agreement across all topologies. The zoomed inset shows experimental data near the classical limit where it is clear that a topological transition from $N\neq0$ to $N=0$ occurs because without entanglement there can be no topology. This robustness of the topological observable is in stark contrast to the decay of typical entanglement witnesses such as concurrence, fidelity and state purity (See SI for the definitions of these quantities), as shown in  Fig.~\ref{Figure 5}(e) using the same example state as that used in Fig.~\ref{Figure 5}(a-c). While each of the entanglement witnesses decay continuously until they each reach their expected minima for a 2D maximally mixed state ($C=0$, $F=0.25$, $\gamma=0.25$), the Skyrmion number remains intact regardless of the degree of noise present, only decaying once the state is maximally mixed.\\

        \begin{figure}
        \includegraphics[width=\linewidth]{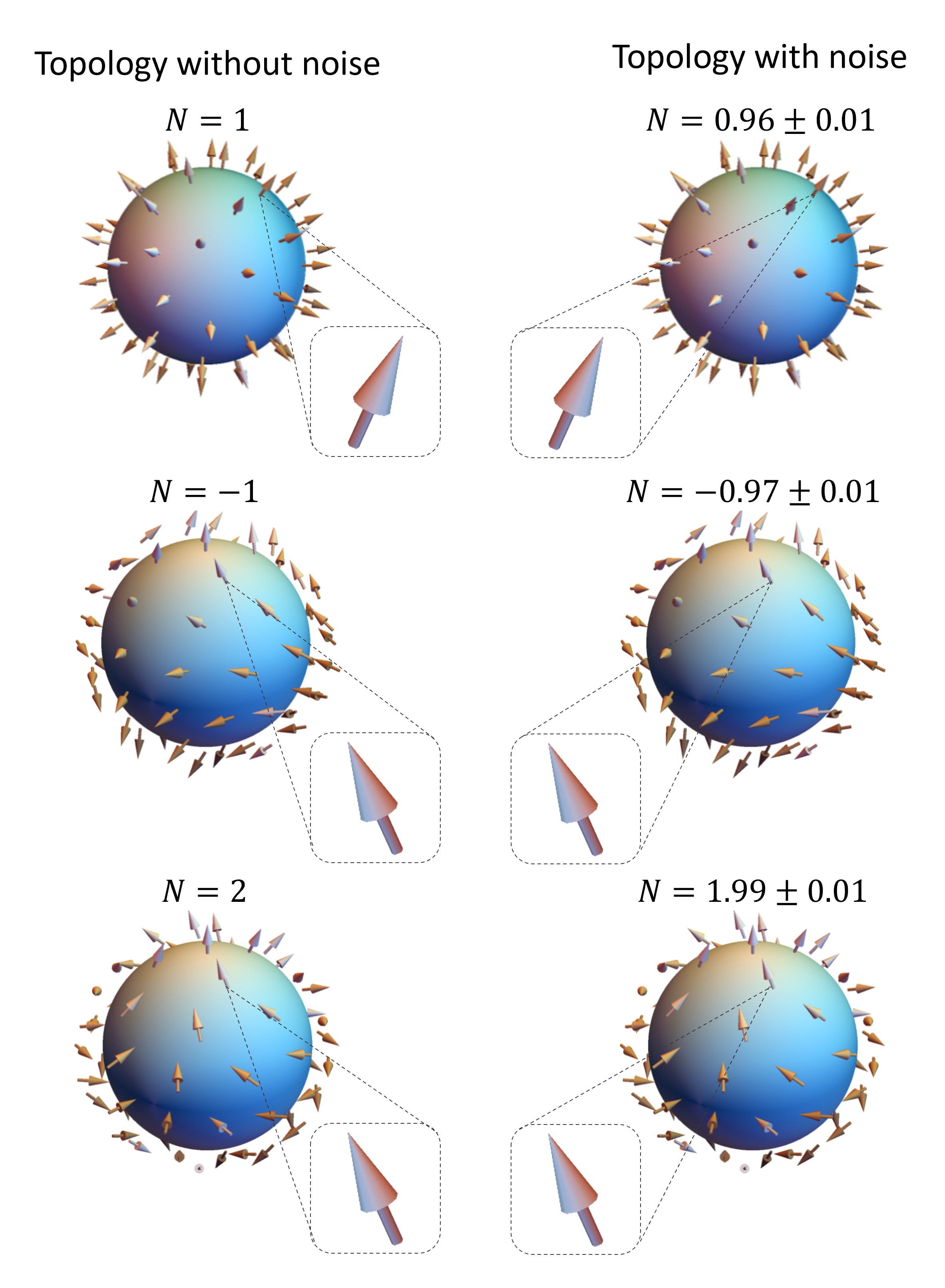}
        \caption{\textbf{Invariance of topology to noise.} The topology of the ideal (Topology without noise) and experimentally retrieved state after passing through the noisy channel (Topology with noise), shown as stereographic projections of the state vectors (normalized to aid visualization) of photon B on a sphere. The calculated Skyrmion number is given for each case, confirming the robustness and the conceptual picture of Fig.~\ref{Figure 2} (b).
        \label{Figure 6}}
    \end{figure}

\section{Discussion}
\noindent 
Typical quantum information encoding schemes rely on the entanglement of non-separable quantum states, a resource that is highly vulnerable to noise, as we have shown. In contrast, we propose encoding quantum information into robust topological observables of entangled states. Despite being rooted in the state's entanglement, these topological features remain entirely intact, even as the state itself degrades when transmitted through a noisy channel. Our approach to noise rejection can be understood in the context of information digitization, where continuous analogue signals are replaced by a discrete signal of bits through ``binning'', e.g., by declaring signals below some threshold as 0 and those above it as 1. At the quantum level, employing discrete subsystems represented as qubits does not achieve the desired discretization. The basic quantum resource, entanglement, is encoded in correlations between the subsystems and simply using subsystems with a finite number of states does not render these correlations discrete. We moot that the topology of the wavefunction with its associated discrete observable naturally achieves this. Why is the discretization of entanglement interesting? Discrete signals are always more robust against the effects of noise. This follows because for discrete signals the noise can't produce an arbitrarily small perturbation of the signal: the noise must be able to flip the signal between two discrete states before any effect is registered. In the same way that digital signals enabled successful classical computation and communication, digital quantum signals may allow successful quantum computation and communication. Given a discretization scheme, it then becomes urgent to establish that noise is not able to drive the jumps between discrete states that would corrupt the discrete signal. The importance of our work is that, for a physical and realistic source of noise, we have been able to demonstrate this. 

Skyrmionic photonic topologies have been celebrated for their potential to act as resilient information carriers \cite{shen2024optical}, but numerical studies have returned mixed outcomes \cite{liu2022disorder}. This report offers, to the best of our knowledge, the first experimental results on their resilience to noise, confirming that such quantum channels are map preserving: they can be viewed as a smooth deformation of the state (map) and thus preserves topology even when the entanglement itself is decaying.  We believe that this noise rejection by topology mechanism could be exploited in practical quantum information processing and communication protocols, e.g., quantum communication under ambient (daylight) conditions, or quantum computing on noisy photonic chips, and could be used in conjunction with the entanglement preserving strategies mentioned in the introduction for preservation of entanglement \textit{and} information. The demonstrated resilience comes without the need for error mitigation or state purification techniques, thus offering a new approach to quantum information resilience in realistic and non-ideal conditions. 

\section{Conclusion}
In conclusion, we have shown both theoretically and experimentally that quantum skyrmions and their topological quantum observables remain resilient to isotropic noise, a generic form of quantum noise that captures realistic effects from the source, channel, and detector, e.g., multi-photon events, dark counts, shot and thermal noise, background light (e.g., daylight) and lost photons.  We present experimental results across various topologies (up to orders $N = \pm 3$) and for a wide range of noise levels, confirming the topological invariance and only collapsing once there is no entanglement at all.  We show that traditional quantum measures decrease continuously under the same conditions, highlighting the advantage of discrete rather than continuous observables.  Our work opens a new path to quantum information processing and communication in noisy quantum systems and channels without the need for compensation or mitigation strategies.

\section*{Acknowledgements}
This work was supported by the South African National Research Foundation/CSIR Rental Pool Programme and the South African Quantum Technology Initiative.

\section*{Author contributions}
The experiment was performed by P.O. and I.N. performed the experiment, and P.O., I.N. and R.M.K. contributed the theory.  All authors contributed to the writing of the manuscript and analysis of data. A.F. conceived of the idea and supervised the project.

\section*{Competing Interests}
The authors declare no competing interests.

\section*{Data availability}
Data will be made available on request.

\renewcommand{\theequation}{S.\arabic{equation}}

\section{Supplementary: Formulation of non-local quantum skyrmion}
\begin{figure}
        \includegraphics[width=\linewidth]{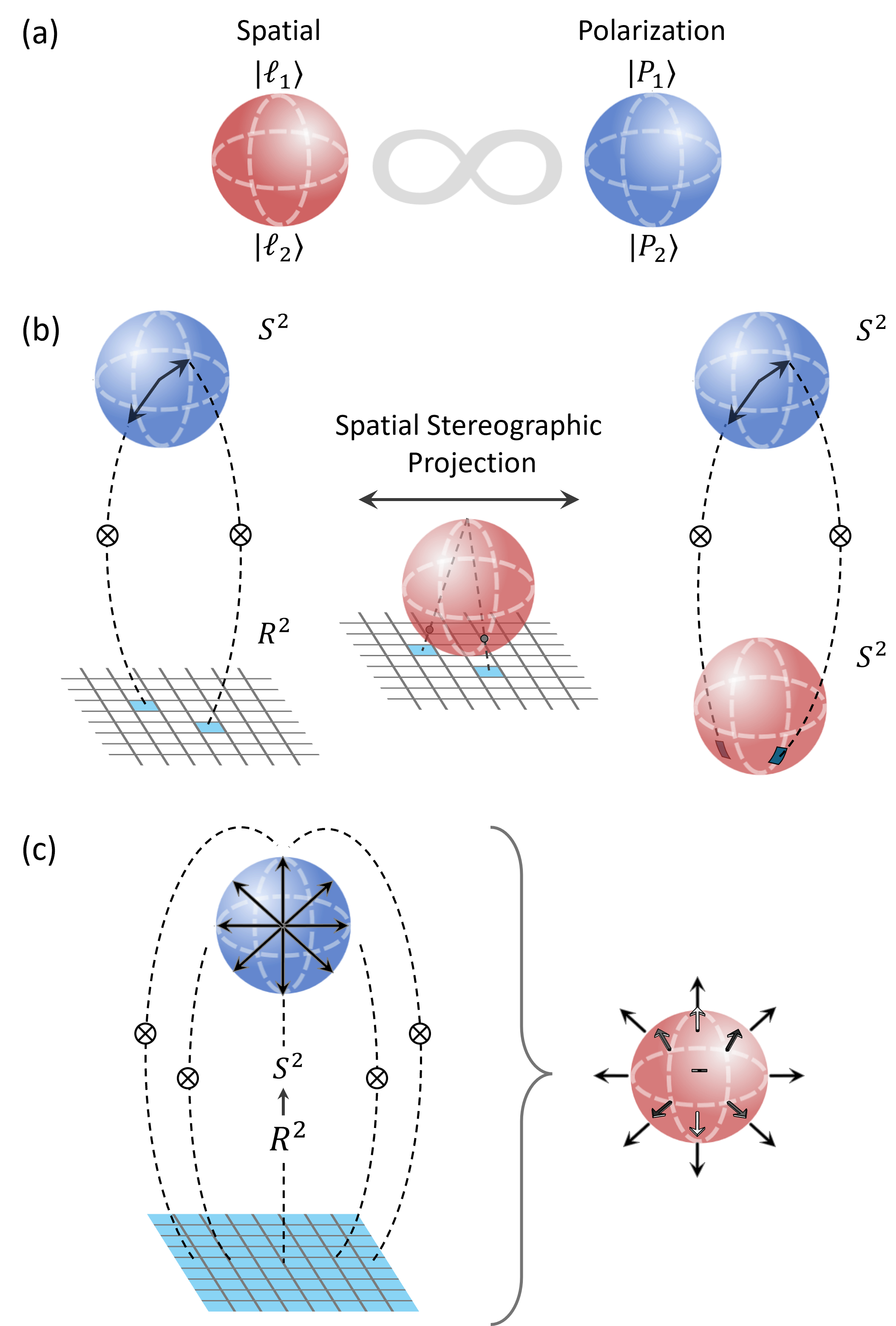}
        \caption{(a) Hybrid entangled state of photons A and B which share OAM-polarization correlations. (b) Such a state also yields position-polarization correlations where a position measurement on photon A yields in coincidence a polarization state collapse for photon B. Through a stereographic projection, this can also be seen as a mapping between a spatial sphere and a polarization sphere. (c) Observation of photon B's state in coincidence with spatial measurements performed on photon A reveals the non-local skyrmionic structure embedded within the entangled state. The topological structure can be compactly represented by combining the spatial sphere of photon A and the polarization state vectors of photon B whose tail is adjacent to its correlated position in photon A.}  
        \label{fig: QSkyStereo} 
    \end{figure}

\noindent A pure non-local biphoton hybrid entangled state of two photons, A and B, correlated in OAM-Polarization (as depicted in Fig.~\ref{fig: QSkyStereo} (a)), can in general be written in the form
    \begin{equation}
        \ket{\Psi} = \frac{1}{\sqrt{2}} \left( \ket{\ell_1}_A \ket{P_1}_B + e^{i\delta}  \ket{\ell_2}_A \ket{P_2}_B \right),
        \label{eq: PureState}
    \end{equation}
where $\ell_1$ and $\ell_2$ denote OAM of $\ell_1 \hbar$ and $\ell_2 \hbar$ per photon, respectively, and $\ket{P_1}, \ket{P_2}$ are orthogonal polarization states, while $\delta$ allows for a relative phase between the two components of the state vector. Photon A's states live on the Hilbert space ${\cal H}_A$ spanned by the states $\{\ket{\ell_1}, \ket{\ell_2} \}$ and photon B's states live on the Hilbert space ${\cal H}_B$ spanned by the states $\{\ket{P_1}, \ket{P_2} \}$. $|\Psi\rangle$ belongs to the Hilbert space ${\cal H}={\cal H}_A\otimes{\cal H}_B$. For states with $|\ell_1| \neq |\ell_2|$ it has been shown that the correlations between these photons form the desired skyrmionic mapping from $\mathcal{S}^2 \to \mathcal{S}^2$ (equivalent to the mapping $\mathcal{R}^2 \to \mathcal{S}^2$ through a stereographic projection of the spatial $\mathcal{S}^2$ to the real plane $\mathcal{R}^2$) as depicted in Fig.~\ref{fig: QSkyStereo} (b) \cite{ornelas2024non}. Expressing photon A in position basis, using $|\ell\rangle = \int_{\mathcal{R}^2} |\text{LG}_\ell \left( \mathbf{r} \right)| e^{i\ell\phi} |\mathbf{r}\rangle d^2r$ where $\text{LG}_\ell \left( \mathbf{r} \right)$ are the Laguerre-Gaussian fields and $|\mathbf{r}\rangle$ are position states, we find
\begin{align}
    \ket{\Psi} = \int_{ \mathcal{R}^2 }  \ket{\mathbf{r}}_A \left( a(\mathbf{r}_A) \ket{P_1}_B + b(\mathbf{r}_A) \ket{P_2}_B \right) d^2r_A,  \label{Eq:SpatialquantumSkyrmion}
\end{align}
    where $a(\mathbf{r}_A) \equiv |\text{LG}_{\ell_{1}} \left( \mathbf{r}_A \right)|$, $b(\mathbf{r}_A) \equiv e^{ i\Theta(\phi_A)}|\text{LG}_{\ell_{2}} \left( \mathbf{r}_A \right)|$, $\Theta(\phi_A) = \Delta\ell\phi_A + \delta$ and $\Delta\ell = \ell_2-\ell_1$. In this formulation, it can be deduced that a spatial measurement on photon A, collapses photon B into a particular polarization state as shown in Fig.~\ref{fig: QSkyStereo} (b). For a non-trivial skyrmionic topology, we have that a full set of spatial measurements on photon A collapses photon B into every possible polarization state as depicted in Fig.~\ref{fig: QSkyStereo} (c). This non-local topology can be depicted compactly by combining the spatial sphere and polarization states together as shown in Fig.~\ref{fig: QSkyStereo} (c) where the tail of each polarization state vector for photon B is adjacent to its correlated position in photon A.
    
    The Skyrmion number of these states can then be calculated using the equation $N = \frac{1}{4\pi}\int_{\mathcal{R}^2} \Sigma_z (x,y) dx dy$ \cite{gao2020paraxial} where $\Sigma_z (x,y)= \epsilon_{pqr} S_p \frac{\partial S_q}{\partial x} \frac{\partial S_r}{\partial y}$ and $\epsilon_{pqr}$ is the Levi-Civita symbol. The integral in the Skyrmion number calculation computes the surface area of photon B's parameter space covered by the mapping defined by the wavefunction given in Eqn.~\ref{Eq:SpatialquantumSkyrmion}.  Thus, since the parameter space is a unit sphere, dividing by the surface area of the sphere ($4\pi$), we find the wrapping number $N$ of the non-local topology. 

The quantum Stokes parameters, $S_i$, are given by the expectation values of the Pauli matrices, which are calculated by taking the diagonal matrix element at position $\mathbf{r}$ for photon A and the partial trace over photon B, such that $S_i = \text{Tr}_B(\langle |\mathbf{r}\rangle_A \,{}_A\langle \mathbf{r}| \otimes \sigma_{B,i} \rangle) = \text{Tr}_B \left(\sigma_{B,i}\,\,{}_A\langle\mathbf{r}|\Psi\rangle \langle \Psi|\mathbf{r}\rangle_A \right)$. From this it can be shown that the Skyrmion number depends on the difference between $\ell_1$ and $\ell_2$, according to $N=m\Delta \ell$ with $m=\text{sign}\left(|\ell_1| - |\ell_2|\right)$. \cite{gao2020paraxial,kuratsuji2021evolution,shen2022generation,shen2021topological} We note that the state could have been expressed in a different basis, such as momentum. However, this would incur a significant computational penalty, as the Fourier transforms required implies that our Skyrmion number calculation would become a complicated double convolution of the Fourier transform of the Stokes parameter. Therefore the continuous position basis is not only more intuitive, it is computationally less cumbersome.

\section{Supplementary: Quantum topological invariance to isotropic noise}
We now consider subjecting our state, $|\Psi\rangle$, to environmental/ ``white" noise, according to the isotropic model where the purity of the state is degraded by mixing it with a maximally mixed state according to  
    \begin{equation}
        \rho = p\ket{\Psi}\bra{\Psi} + \frac{1-p}{d^2} \mathbb{1}_{d^2}, 
        \label{eq: PartialMixed}
    \end{equation}

\noindent
    where $d=2$, $p \in [0,1]$ is a parameter controlling the degree of purity of the state with $p=1$ giving a pure state ($|\Psi\rangle \langle \Psi|$) and $p=0$ giving a maximally mixed state ($\frac{1}{d^2}\mathbb{1}_{d^2}$) and $\mathbb{1}_{d^2}$, the $d^2 \,\times\,d^2$ identity matrix, is the identity operator on ${\cal H}$. The purity, $\gamma$, of a state is given by $\text{Tr}(\rho^2)$. It follows that $\gamma$ is related to $p$ according to 
\begin{eqnarray}
\gamma &=& \text{Tr}\left(\left(p\ket{\Psi}\bra{\Psi} + \frac{1-p}{d^2} \mathbb{1}_{d^2}\right)^2\right) \nonumber \\
&=& \text{Tr}\left(p^2\ket{\Psi}\bra{\Psi} + 2p\frac{(1-p)}{d^2}\ket{\Psi}\bra{\Psi} + \frac{(1-p)^2}{d^4} \mathbb{1}_{d^2}\right) \nonumber\\
&=& p^2 + 2p\frac{(1-p)}{d^2} + \frac{(1-p)^2}{d^2} \nonumber \\
&=& p^2 + \frac{1-p^2}{d^2}.
\label{eq: purity_p}
\end{eqnarray}

    \noindent To calculate the Skyrmion number we must consider how the quantum Stokes parameters are affected by the isotropic noise with the model given above,
    \begin{eqnarray}
        S_i' &=&\text{Tr}_B \left(\sigma_{B,i}\,{}_A\langle\mathbf{r}|\rho|\mathbf{r}\rangle_A\right)\cr\cr
        &=&\text{Tr}_B \Big(\sigma_{B,i}\big(p\,{}_A\langle\mathbf{r}|\Psi\rangle\langle\Psi|\mathbf{r}\rangle_A\cr\cr
        &&\qquad+\frac{1-p}{4}\,{}_A\langle\mathbf{r}|\mathbb{1}_4|\mathbf{r}\rangle_A\big) \Big)
        \label{eq: PartialMixedPauli}
    \end{eqnarray}
t should be noted that without loss of generality we have chosen to restrict ourselves to the original 2D OAM hilbert space of photon A, by projecting onto only that hilbert space. This is because isotropic noise treats every OAM state in the exact same way (mixing it with the identity), thus we are justified in only considering states within which we start with a non-zero signal contribution. To compute the diagonal matrix element of $\mathbb{1}_4$, start by noting that we have
\begin{equation}
a(\mathbf{r})\,\,=\,\, \langle\mathbf{r}|\ell_1\rangle\qquad\qquad
b(\mathbf{r})\,\,=\,\, \langle\mathbf{r}|\ell_2\rangle
\label{overlaps}
\end{equation}
which follow by comparing (\ref{eq: PureState}) and (\ref{Eq:SpatialquantumSkyrmion}). We note that we have ignored the factor $\frac{1}{\sqrt{2}}$ in (\ref{overlaps}) since to obtain correctly normalized quantum Stokes parameters, we must choose
\begin{equation}
|a|^2+|b|^2=1\label{norms}
\end{equation}
at all $\mathbf{r}$. Using
\begin{eqnarray}
\mathbb{1}_4&=&|\ell_1\rangle_A|P_1\rangle_B \,{}_A\langle\ell_1|\,{}_B\langle P_1|
+|\ell_1\rangle_A|P_2\rangle_B \,{}_A\langle\ell_1|\,{}_B\langle P_2|\cr\cr
&&+|\ell_2\rangle_A|P_1\rangle_B \,{}_A\langle\ell_2|\,{}_B\langle P_1|+
|\ell_2\rangle_A|P_2\rangle_B \,{}_A\langle\ell_2|\,{}_B\langle P_2|
\nonumber
\end{eqnarray}
we easily find (use (\ref{overlaps}) and (\ref{norms}))
\begin{eqnarray}
\langle\mathbf{r}|\mathbb{1}_4|\mathbf{r}\rangle&=&2|a|^2|P_1\rangle_B \,{}_B\langle P_1|
+2|a|^2|P_2\rangle_B \,{}_B\langle P_2|\cr\cr
&&+2|b|^2|P_1\rangle_B \,{}_B\langle P_1|+
2|b|^2|P_2\rangle_B \,{}_B\langle P_2|\cr\cr
&=& 2\mathbb{1}_2
\nonumber
\end{eqnarray}
where $\mathbb{1}_2$ is the identity operator on ${\cal H}_B$ and $a \equiv a(\mathbf{r})$ and   $ b \equiv b(\mathbf{r})$ . Consequently we have
    \begin{eqnarray}
        S_i' &=&\text{Tr}_B\Big(\sigma_{B,i}\big(p\,{}_A\langle\mathbf{r}|\Psi\rangle\langle\Psi|\mathbf{r}\rangle_A+\frac{1-p}{2}\mathbb{1}_2\big)\Big)
        \label{eq: PartialMixedPauli}
    \end{eqnarray}
    
    
    \noindent It is satisfying that after taking the diagonal matrix element, the maximally mixed state in ${\cal H}$ has been replaced by the maximally mixed state in ${\cal H}_B$. Using the linearity of the trace operation and the fact that the Pauli matrices are traceless, we have
    \begin{eqnarray}
        S_i'&=&p\text{Tr}_B\left( \sigma_{B,i}\,\,{}_A\langle\mathbf{r}|\Psi\rangle\langle\Psi|\mathbf{r}\rangle_A \right)\cr\cr
        &&+\frac{1-p}{2}\text{Tr}_B\left(\sigma_{B,i} \right)\cr\cr
        &=&p\,\text{Tr}_B\left( \sigma_{B,i}\,\,{}_A\langle\mathbf{r}|\Psi\rangle\langle\Psi|\mathbf{r}\rangle_A \right)\cr\cr
        &=&p S_i
         \label{eq: PartialMixedPauliResult}
    \end{eqnarray}

    \noindent where $S_i$ are the quantum Stokes parameters for the pure state without isotropic noise. Therefore, the effect of isotropic noise on the state is to multiply the Stokes parameters, $S_i$, by a constant factor $p$. The Skyrmion number assumes normalization such that $\Sigma_i^3 S_i^2 = 1$. After correct normalization of the Stokes parameters this constant $p$ factor does not contribute demonstrating that isotropic noise does not alter the topology. This demonstrates that the skyrmionic topology of the state is invariant to isotropic noise as long as a portion of the pure state survives i.e., $p>0$ or equivalently $\gamma>\frac{1}{d^2}$. 
    
    \section{Supplementary: Rejection mechanism underlying projective measurements}
    
    To further clarify the result obtained in the previous section, we will now argue that the invariance of the skyrmionic topology to isotropic noise can be understood as noise rejection by projective measurements made on our state. Towards this end we employ the spectral decomposition of the Pauli matrices, $\sigma_{B,i}= \lambda^+_iP^+_i + \lambda^-_iP^-_i$ where $P_i^{\pm} = |\lambda^{\pm}_i\rangle\langle\lambda^{\pm}_i|$ and the eigenvalues are $\lambda^{\pm}_i=\pm 1$. Substituting this into Eqn. \ref{eq: PartialMixedPauliResult} we find
    \begin{eqnarray}
        S_i'&=&\text{Tr}_B \left(P^+_i \left[p\,\,{}_A\langle\mathbf{r}|\Psi\rangle\langle\Psi|\mathbf{r}\rangle_A + \frac{1-p}{2}\mathbb{1_{2}} \right]\right)\cr\cr\cr
        &&-\text{Tr}_B \left(P^-_i \left[p\,\,{}_A\langle\mathbf{r}|\Psi\rangle\langle\Psi|\mathbf{r}\rangle_A + \frac{1-p}{2}\mathbb{1_{2}} \right]\right)\cr\cr
        &=&(I_{i,\text{pure}}^+ + I^+_{i,\text{noise}}) - (I_{i,\text{pure}}^- + I^-_{i,\text{noise}}),\cr
        &&
    \end{eqnarray}
    
    where $I^{\pm}_{i,\text{pure}}=\text{Tr}_B \left(P^{\pm}_i p\,\,{}_A\langle\mathbf{r}|\Psi\rangle\langle\Psi|\mathbf{r}\rangle_A\right)$ and $I^{\pm}_{i,\text{noise}} = \text{Tr}_B \left(P^{\pm}_i \frac{1-p}{2}\mathbb{1_2} \right)$. Experimentally, we measure the quantities $I_{i,\text{exp}}^+ = I_{i,\text{pure}}^+ + I^+_{i,\text{noise}}$ and $I_{i,\text{exp}}^- = I_{i,\text{pure}}^- + I^-_{i,\text{noise}}$. Since the eigenvalues of the Pauli matrices are non-degenerate we have that $\text{Tr}\left(P^+_i\right) = \text{Tr}\left(P^-_i\right)=1$. Therefore $I^{+}_{i,\text{noise}} = I^{-}_{i,\text{noise}}$, so that $S_i' = I_{i,\text{exp}}^+ - I_{i,\text{exp}}^- = I_{i,\text{pure}}^+ - I_{i,\text{pure}}^+$. This shows that each pair of projective measurements required to calculate each Pauli observable, receive identical noise contributions, which thus cancels in their difference, explaining the noise rejection. 

\section{Supplementary: Experiment}

\begin{figure*}[t!]
        \includegraphics[width=\textwidth]{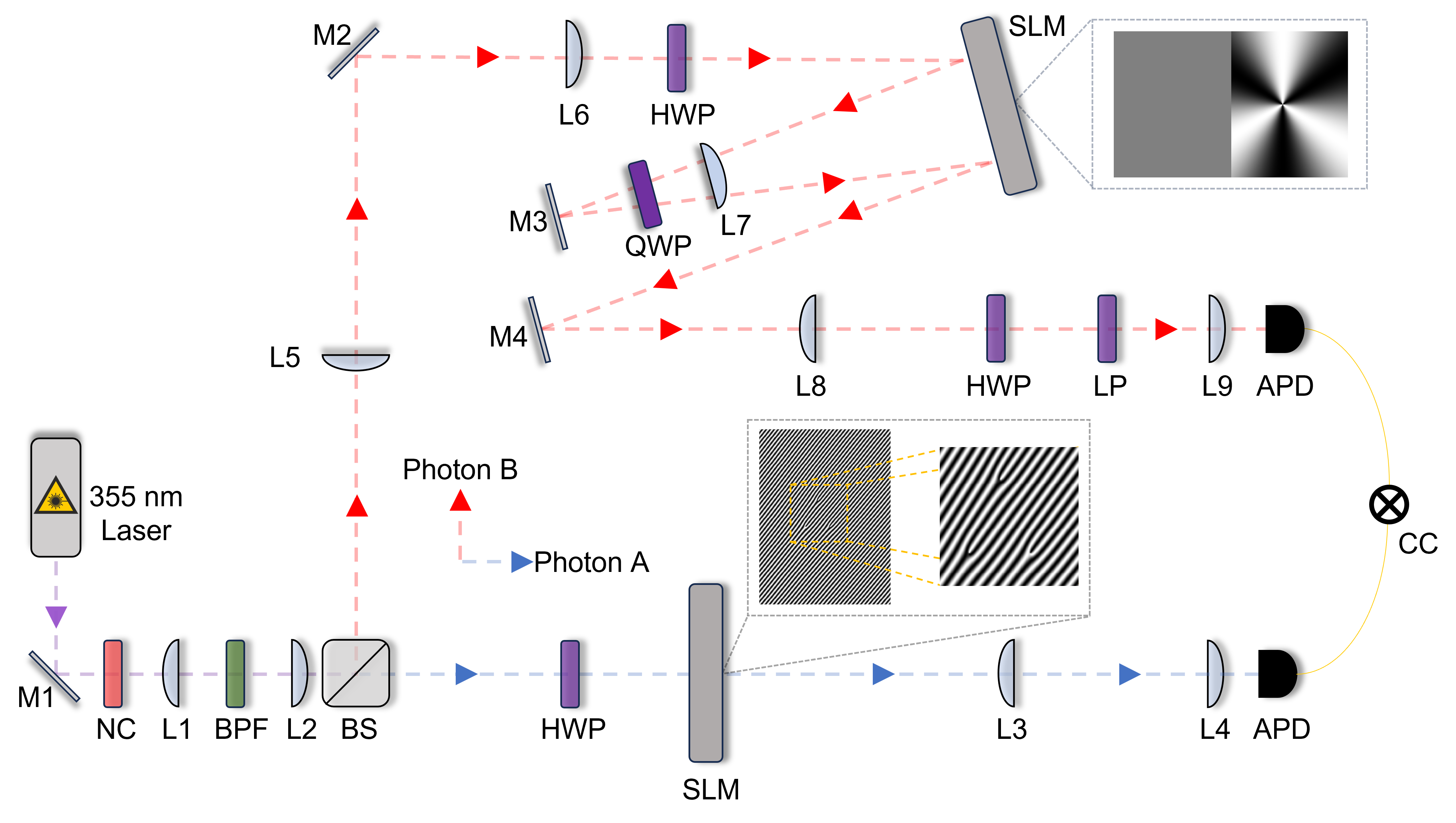}
        \caption{Experimental setup for the generation and detection of non-local skyrmionic states. Some abbreviations:  mirror (M), non-linear crystal (NC), lens (L), band-pass filter (BPF), 50:50 beam splitter (BS), half-wave plate (HWP), spatial light modulator (SLM), quarter-wave plate (QWP), linear polarizer (LP), avalanche photodiode (APD), coincidence counter (CC).}
        \label{fig:ExpFig}
    \end{figure*}

\noindent A schematic of the experiment conducted is shown in Fig.~\ref{fig:ExpFig}. Entangled photon pairs were generated through spontaneous parametric down-conversion (SPDC), where a 355 nm wavelength, collimated Gaussian beam was sent through a 3 mm long, Type-I Barium Borate (BBO) non-linear crystal (NC). A band-pass filter (BPF) centred at 710 nm wavelength was used to filter out the unconverted pump beam. From the SPDC process, the generated photons were correlated in OAM, sharing the non-separable state, $|\Psi\rangle = \Sigma_{\ell} c_{\ell} |\ell\rangle_A \otimes |-\ell\rangle_B$, where the coefficients, $c_{\ell}$, determined the weightings for each subspace that was spanned by the OAM eigenstates, $\ket{ \pm \ell}$, for each photon. The two entangled photons (photon A and B) were then spatially separated using a 50:50 beam-splitter (BS). In order to prepare the desired non-local skyrmionic state the initial OAM-OAM entangled state was mapped to an arbitrary hybrid entangled state of the form

\begin{equation}
    | \Psi \rangle = \frac{1}{\sqrt{2}}\left(|\ell_1\rangle_A |H\rangle_B + |\ell_2\rangle_A |V\rangle_B\right), 
    \label{eq: suppHybEnt}
\end{equation}

through the use of a digital spatial-to-polarization coupling (SPC) approach \cite{Nape_2022} where the spatial information of photon B was coupled to polarization. To achieve the desired arbitrary state given in Eq.~\ref{eq: suppHybEnt}, the OAM DOF of photon B was coupled to orthogonal polarization states using a post-selection of the desired OAM subspace with separate modulations by an SLM. 
The transformations undergone by the state of photon B due to the entire SPC can be broken down as follows (we ignore contributions from states which will not couple into the fibre, i.e we only consider $\ket{m} = \ket{0}$) $\ket{H,\ell^{'}_1} + \ket{V,\ell^{'}_2} \xrightarrow{\text{SLM}} \ket{H,\ell^{'}_1+\ell_1} + \ket{V,\ell^{'}_2} \xrightarrow{\text{QWP}} \ket{L,\ell^{'}_1+\ell_1} + \ket{R,\ell^{'}_2} \xrightarrow{\text{M}} \ket{R,\ell^{'}_1+\ell_1} + \ket{L,\ell^{'}_2} \xrightarrow{\text{QWP}} \ket{V,\ell^{'}_1+\ell_1} + \ket{H,\ell^{'}_2} \xrightarrow{\text{SLM}} \ket{V,\ell^{'}_1+\ell_1} + \ket{H,\ell^{'}_2 + \ell_2}$, where coupling into the fibre ensures that $\ell_1 = -\ell^{'}_1$ and $\ell_2 = -\ell^{'}_2$ thereby erasing the OAM information from photon B as it becomes a separable DOF in photon B. It is clear then that the desired OAM subspace, $\{\ell_1,\ell_2\}$ is selected by displaying those modes on the SLM. Following the state preparation, spatially separated projective measurements were performed on both photons and they were observed in coincidence allowing for the construction of a full quantum state tomography of the biphoton state. Photon A was detected through a coupled detection system consisting of an SLM and SMF coupled to an avalanche photon detector (APD). Photon B was measured using a set of polarization optics, a HWP orientated to $45^{\circ}$ and a linear polarizer (LP) orientated at $90^{\circ}$.  

\section{Supplementary: Purity and Quantum contrast}

The purity of the states generated within the experiment can directly be controlled by controlling the quantum contrast. The discussion that follows is based on the supplementary material given in reference \cite{zhu2021high}.

Using the result given by Eq.~\ref{eq: purity_p} it is clear that the purity of a state is dependent on the ratio between the signal and isotropic noise component of the state, $p$. This can directly be attributed to the quantum contrast $Q_c$ of the state, which gives the ratio between the accidental and two-photon coincidences. This allows a relation between the probability $p$ and $Q_c$ as follows
\begin{align}
    Q_c = \frac{1-p+pd}{1-p} \implies p = \frac{Q_c-1}{Q_c-1+d}
\end{align}
where $d$ is the dimensionality of the state. Substituting Eq.~\ref{eq: purity_p} and solving for $\gamma$ then gives 
\begin{equation}
\gamma=\frac{d\big(Q_c^2-2Q_c+2\big)+2 (Q_c-1)}{d (d+Q_c-1)^2} \label{eq: PurityQc_supp}.
\end{equation}
Experimentally a full quantum state tomography is performed, consisting of 36 projective measurements (for $d=2$) each with its own $Q_c$. When computing the $Q_c$ for a state, the average $Q_c$ across all 36 measurements is used. Furthermore, the $Q_c$ is directly computed from the accidental coincidences and two photon coincidences according to
\begin{align}
    Q_c = \frac{1}{T}\frac{C}{AB}
\end{align}   
where $C$ is the two photon coincidences and $N_{acc} = TAB$ is the accidental coincidences with $T$ being the coincidence window and $A,B$ being the number of single photon detection events detected by the APDs in the optical path of photon A and B, respectively. With this in mind, the $Q_c$ value used in Eq.~\ref{eq: PurityQc_supp} is given by
\begin{align}
    Q_c = \frac{1}{36T}\sum_{i,j=1}^6 \frac{C_{i.j}}{A_{i.j}B_{i.j}}
\end{align}   
where the indices $i,j$ reference a particular projective measurement used to build the full QST. 

\section{Supplementary: Concurrence and Fidelity}

\noindent Beyond purity, quantities such as Concurrence and Fidelity also serve as entanglement witnesses for our states.\\

The fidelity was used to analytically compare our measured partially mixed states, $\rho$ against an initial pure state $\rho_T=|\Psi\rangle\langle\Psi|$
\begin{equation}
    F =\left( \text{Tr}  \left( \sqrt{  \sqrt{\rho_T}\rho \sqrt{\rho_T}  }  \right) \right)^2 ,
\end{equation}
 The fidelity is 0 if the states are not identical or 1 when they are identical up to a global phase. However, since $\rho$ is a partially mixed state, we find that when $\rho$ becomes completely mixed, that is $\rho = \frac{1}{4} \mathbb{1}_{d^2}$, then $F=\frac{1}{4}$.\\

The concurrence was used to measure the degree of entanglement between the hybrid entangled photons. It was measured from

\begin{equation}
    C(\rho) = \text{max} \{ 0, \lambda_1 -\lambda_2- \lambda_3 - \lambda_4 \},
\end{equation}

where $\lambda_i$ are eigenvalues of the operator $ R = \text{Tr} \left( \sqrt{  \sqrt{\rho} \tilde{\rho} \sqrt{\rho}  }  \right)$ in descending order and $\tilde{\rho} = \sigma_{y} \otimes \sigma_{y} \rho^* \sigma_{y} \otimes \sigma_{y}$. The concurrence ranges from 0 for separable and completely mixed states to 1 for entangled states.


\begin{thebibliography}{10}

\bibitem{gisin2002quantum}
N.~Gisin, G.~Ribordy, W.~Tittel, and H.~Zbinden, ``Quantum cryptography,'' {\em Reviews of Modern Physics}, vol.~74, no.~1, p.~145, 2002.

\bibitem{scarani2009security}
V.~Scarani, H.~Bechmann-Pasquinucci, N.~J. Cerf, M.~Du{\v{s}}ek, N.~L{\"u}tkenhaus, and M.~Peev, ``The security of practical quantum key distribution,'' {\em Reviews of Modern Physics}, vol.~81, no.~3, p.~1301, 2009.

\bibitem{ma2007quantum}
X.~Ma, C.-H.~F. Fung, and H.-K. Lo, ``Quantum key distribution with entangled photon sources,'' {\em Physical Review A}, vol.~76, no.~1, p.~012307, 2007.

\bibitem{liao2017satellite}
S.-K. Liao, W.-Q. Cai, W.-Y. Liu, L.~Zhang, Y.~Li, J.-G. Ren, J.~Yin, Q.~Shen, Y.~Cao, Z.-P. Li, {\em et~al.}, ``Satellite-to-ground quantum key distribution,'' {\em Nature}, vol.~549, no.~7670, pp.~43--47, 2017.

\bibitem{bedington2017progress}
R.~Bedington, J.~M. Arrazola, and A.~Ling, ``Progress in satellite quantum key distribution,'' {\em npj Quantum Information}, vol.~3, no.~1, p.~30, 2017.

\bibitem{thomas2022efficient}
P.~Thomas, L.~Ruscio, O.~Morin, and G.~Rempe, ``Efficient generation of entangled multiphoton graph states from a single atom,'' {\em Nature}, vol.~608, no.~7924, pp.~677--681, 2022.

\bibitem{raussendorf2001one}
R.~Raussendorf and H.~J. Briegel, ``A one-way quantum computer,'' {\em Physical Review Letters}, vol.~86, no.~22, p.~5188, 2001.

\bibitem{briegel2009measurement}
H.~J. Briegel, D.~E. Browne, W.~D{\"u}r, R.~Raussendorf, and M.~Van~den Nest, ``Measurement-based quantum computation,'' {\em Nature Physics}, vol.~5, no.~1, pp.~19--26, 2009.

\bibitem{pepe2016correlation}
F.~V. Pepe, F.~Di~Lena, A.~Garuccio, G.~Scarcelli, and M.~D’Angelo, ``Correlation plenoptic imaging with entangled photons,'' {\em Technologies}, vol.~4, no.~2, p.~17, 2016.

\bibitem{sephton2023revealing}
B.~Sephton, I.~Nape, C.~Moodley, J.~Francis, and A.~Forbes, ``Revealing the embedded phase in single-pixel quantum ghost imaging,'' {\em Optica}, vol.~10, no.~2, pp.~286--291, 2023.

\bibitem{giovannetti2011advances}
V.~Giovannetti, S.~Lloyd, and L.~Maccone, ``Advances in quantum metrology,'' {\em Nature photonics}, vol.~5, no.~4, pp.~222--229, 2011.

\bibitem{boto2000quantum}
A.~N. Boto, P.~Kok, D.~S. Abrams, S.~L. Braunstein, C.~P. Williams, and J.~P. Dowling, ``Quantum interferometric optical lithography: exploiting entanglement to beat the diffraction limit,'' {\em Physical Review Letters}, vol.~85, no.~13, p.~2733, 2000.

\bibitem{kok2001quantum}
P.~Kok, A.~N. Boto, D.~S. Abrams, C.~P. Williams, S.~L. Braunstein, and J.~P. Dowling, ``Quantum-interferometric optical lithography: Towards arbitrary two-dimensional patterns,'' {\em Physical Review A}, vol.~63, no.~6, p.~063407, 2001.

\bibitem{nielsen2001quantum}
M.~A. Nielsen and I.~L. Chuang, {\em Quantum computation and quantum information}, vol.~2.
\newblock Cambridge university press Cambridge, 2001.

\bibitem{ecker2019overcoming}
S.~Ecker, F.~Bouchard, L.~Bulla, F.~Brandt, O.~Kohout, F.~Steinlechner, R.~Fickler, M.~Malik, Y.~Guryanova, R.~Ursin, {\em et~al.}, ``Overcoming noise in entanglement distribution,'' {\em Physical Review X}, vol.~9, no.~4, p.~041042, 2019.

\bibitem{zhu2021high}
F.~Zhu, M.~Tyler, N.~H. Valencia, M.~Malik, and J.~Leach, ``Is high-dimensional photonic entanglement robust to noise?,'' {\em AVS Quantum Science}, vol.~3, no.~1, p.~011401, 2021.

\bibitem{kumar2003effect}
D.~Kumar and P.~Pandey, ``Effect of noise on quantum teleportation,'' {\em Physical Review A}, vol.~68, no.~1, p.~012317, 2003.

\bibitem{lloyd1997capacity}
S.~Lloyd, ``Capacity of the noisy quantum channel,'' {\em Physical Review A}, vol.~55, no.~3, p.~1613, 1997.

\bibitem{liang2013quantum}
H.-Q. Liang, J.-M. Liu, S.-S. Feng, and J.-G. Chen, ``Quantum teleportation with partially entangled states via noisy channels,'' {\em Quantum Information Processing}, vol.~12, no.~8, pp.~2671--2687, 2013.

\bibitem{peterfreund2021multidimensional}
E.~Peterfreund and M.~Gavish, ``Multidimensional scaling of noisy high dimensional data,'' {\em Applied and Computational Harmonic Analysis}, vol.~51, pp.~333--373, 2021.

\bibitem{almeida2007noise}
M.~L. Almeida, S.~Pironio, J.~Barrett, G.~T{\'o}th, and A.~Ac{\'\i}n, ``Noise robustness of the nonlocality of entangled quantum states,'' {\em Physical Review Letters}, vol.~99, no.~4, p.~040403, 2007.

\bibitem{qu2022robust}
R.~Qu, Y.~Wang, X.~Zhang, S.~Ru, F.~Wang, H.~Gao, F.~Li, and P.~Zhang, ``Robust method for certifying genuine high-dimensional quantum steering with multimeasurement settings,'' {\em Optica}, vol.~9, no.~5, pp.~473--478, 2022.

\bibitem{tsokeng2018dynamics}
A.~T. Tsokeng, M.~Tchoffo, and L.~C. Fai, ``Dynamics of entanglement and quantum states transitions in spin-qutrit systems under classical dephasing and the relevance of the initial state,'' {\em Journal of Physics Communications}, vol.~2, no.~3, p.~035031, 2018.

\bibitem{qu2022retrieving}
R.~Qu, Y.~Wang, M.~An, F.~Wang, Q.~Quan, H.~Li, H.~Gao, F.~Li, and P.~Zhang, ``Retrieving high-dimensional quantum steering from a noisy environment with n measurement settings,'' {\em Physical Review Letters}, vol.~128, no.~24, p.~240402, 2022.

\bibitem{nape2023quantum}
I.~Nape, B.~Sephton, P.~Ornelas, C.~Moodley, and A.~Forbes, ``Quantum structured light in high dimensions,'' {\em APL Photonics}, vol.~8, no.~5, 2023.

\bibitem{ndagano2017characterizing}
B.~Ndagano, B.~Perez-Garcia, F.~S. Roux, M.~McLaren, C.~Rosales-Guzman, Y.~Zhang, O.~Mouane, R.~I. Hernandez-Aranda, T.~Konrad, and A.~Forbes, ``Characterizing quantum channels with non-separable states of classical light,'' {\em Nature Physics}, vol.~13, no.~4, pp.~397--402, 2017.

\bibitem{yan2023advances}
P.-S. Yan, L.~Zhou, W.~Zhong, and Y.-B. Sheng, ``Advances in quantum entanglement purification,'' {\em Science China Physics, Mechanics \& Astronomy}, vol.~66, no.~5, p.~250301, 2023.

\bibitem{yan2021quantum}
Q.~Yan, X.~Hu, Y.~Fu, C.~Lu, C.~Fan, Q.~Liu, X.~Feng, Q.~Sun, and Q.~Gong, ``Quantum topological photonics,'' {\em Advanced Optical Materials}, vol.~9, no.~15, p.~2001739, 2021.

\bibitem{mehrabad2020chiral}
M.~J. Mehrabad, A.~P. Foster, R.~Dost, E.~Clarke, P.~K. Patil, A.~M. Fox, M.~S. Skolnick, and L.~R. Wilson, ``Chiral topological photonics with an embedded quantum emitter,'' {\em Optica}, vol.~7, no.~12, pp.~1690--1696, 2020.

\bibitem{dai2022topologically}
T.~Dai, Y.~Ao, J.~Bao, J.~Mao, Y.~Chi, Z.~Fu, Y.~You, X.~Chen, C.~Zhai, B.~Tang, {\em et~al.}, ``Topologically protected quantum entanglement emitters,'' {\em Nature Photonics}, vol.~16, no.~3, pp.~248--257, 2022.

\bibitem{mittal2018topological}
S.~Mittal, E.~A. Goldschmidt, and M.~Hafezi, ``A topological source of quantum light,'' {\em Nature}, vol.~561, no.~7724, pp.~502--506, 2018.

\bibitem{barik2018topological}
S.~Barik, A.~Karasahin, C.~Flower, T.~Cai, H.~Miyake, W.~DeGottardi, M.~Hafezi, and E.~Waks, ``A topological quantum optics interface,'' {\em Science}, vol.~359, no.~6376, pp.~666--668, 2018.

\bibitem{blanco2018topological}
A.~Blanco-Redondo, B.~Bell, D.~Oren, B.~J. Eggleton, and M.~Segev, ``Topological protection of biphoton states,'' {\em Science}, vol.~362, no.~6414, pp.~568--571, 2018.

\bibitem{parmee2022optical}
C.~D. Parmee, M.~R. Dennis, and J.~Ruostekoski, ``Optical excitations of skyrmions, knotted solitons, and defects in atoms,'' {\em Communications Physics}, vol.~5, no.~1, p.~54, 2022.

\bibitem{shen2024optical}
Y.~Shen, Q.~Zhang, P.~Shi, L.~Du, X.~Yuan, and A.~V. Zayats, ``Optical skyrmions and other topological quasiparticles of light,'' {\em Nature Photonics}, vol.~18, no.~1, pp.~15--25, 2024.

\bibitem{gao2020paraxial}
S.~Gao, F.~C. Speirits, F.~Castellucci, S.~Franke-Arnold, S.~M. Barnett, and J.~B. G{\"o}tte, ``Paraxial skyrmionic beams,'' {\em Physical Review A}, vol.~102, no.~5, p.~053513, 2020.

\bibitem{shen2021topological}
Y.~Shen, ``Topological bimeronic beams,'' {\em Optics Letters}, vol.~46, no.~15, pp.~3737--3740, 2021.

\bibitem{shen2022generation}
Y.~Shen, E.~C. Mart{\'\i}nez, and C.~Rosales-Guzm{\'a}n, ``Generation of optical skyrmions with tunable topological textures,'' {\em ACS Photonics}, vol.~9, no.~1, pp.~296--303, 2022.

\bibitem{singh2023synthetic}
K.~Singh, P.~Ornelas, A.~Dudley, and A.~Forbes, ``Synthetic spin dynamics with bessel-gaussian optical skyrmions,'' {\em Optics Express}, vol.~31, no.~10, pp.~15289--15300, 2023.

\bibitem{sugic2021particle}
D.~Sugic, R.~Droop, E.~Otte, D.~Ehrmanntraut, F.~Nori, J.~Ruostekoski, C.~Denz, and M.~R. Dennis, ``Particle-like topologies in light,'' {\em Nature communications}, vol.~12, no.~1, pp.~1--10, 2021.

\bibitem{ornelas2024non}
P.~Ornelas, I.~Nape, R.~de~Mello~Koch, and A.~Forbes, ``Non-local skyrmions as topologically resilient quantum entangled states of light,'' {\em Nature Photonics}, pp.~1--9, 2024.

\bibitem{liu2022disorder}
C.~Liu, S.~Zhang, S.~A. Maier, and H.~Ren, ``Disorder-induced topological state transition in the optical skyrmion family,'' {\em Physical Review Letters}, vol.~129, no.~26, p.~267401, 2022.

\bibitem{horodecki2000limits}
M.~Horodecki, P.~Horodecki, and R.~Horodecki, ``Limits for entanglement measures,'' {\em Physical Review Letters}, vol.~84, no.~9, p.~2014, 2000.

\bibitem{horodecki1999general}
M.~Horodecki, P.~Horodecki, and R.~Horodecki, ``General teleportation channel, singlet fraction, and quasidistillation,'' {\em Physical Review A}, vol.~60, no.~3, p.~1888, 1999.

\bibitem{horodecki2009quantum}
R.~Horodecki, P.~Horodecki, M.~Horodecki, and K.~Horodecki, ``Quantum entanglement,'' {\em Reviews of modern physics}, vol.~81, no.~2, p.~865, 2009.

\bibitem{dur1999quantum}
W.~D{\"u}r, H.-J. Briegel, J.~I. Cirac, and P.~Zoller, ``Quantum repeaters based on entanglement purification,'' {\em Physical Review A}, vol.~59, no.~1, p.~169, 1999.

\bibitem{kuratsuji2021evolution}
H.~Kuratsuji and S.~Tsuchida, ``Evolution of the stokes parameters, polarization singularities, and optical skyrmion,'' {\em Physical Review A}, vol.~103, no.~8, p.~023514, 2021.

\bibitem{Nape_2022}
I.~Nape, A.~G. de~Oliveira, D.~Slabbert, N.~Bornman, J.~Francis, P.~H.~S. Ribeiro, and A.~Forbes, ``An all-digital approach for versatile hybrid entanglement generation,'' {\em Journal of Optics}, vol.~24, p.~054003, mar 2022.

\end{thebibliography}
\end{document}